\documentclass[10pt,referee,iopams,setstack,epsfig,epsf,float]{iopart}
\usepackage{cite}
\usepackage{epsf}
\usepackage{epsfig}
\usepackage{float}
\usepackage{amssymb}

\def\ep{{\epsilon}}

\def\k{{{\bf k}}}
\def\om{{\omega}}
\def\omt{{\tilde{\omega}}}

\def\nnu{{\nonumber}} 

\def\g{{\bf{g}}}

\def\P{{\bf{P}}}

\def\beq{\begin{equation}}
\def\eeq{\end{equation}}
\def\beqa{\begin{eqnarray}}
\def\eeqa{\end{eqnarray}}

\def\g0{{\gamma_0}}

\eqnobysec
\begin{document}
\bibliographystyle{plain}
\input epsf

\title[Interaction effects in mixed-valent Kondo insulators]{ Interaction
effects in mixed-valent Kondo insulators}

\author{Anne Gilbert\dag, N.S.Vidhyadhiraja\ddag, David E Logan\dag }

\address{\dag\ University of Oxford, Physical and Theoretical Chemistry
Laboratory,\\ South Parks Rd, Oxford OX1~3QZ, UK}
\address{
\ddag\ Jawaharlal Nehru Centre for Advanced Scientific Research,\\
Bangalore 560~064, India}

\begin{abstract}
We study theoretically the class of mixed-valent Kondo insulators, employing a 
recently developed local moment approach to heavy Fermion systems using the {\em
asymmetric} periodic Anderson model (PAM). Novel features in spectra and transport, observable experimentally but lying outside the scope of the symmetric PAM or the Kondo lattice model, emerge naturally within the present theory. We argue in particular that a shoulder-like feature in the optical conductivity, that is distinct from the usual 
mid-infrared or direct gap peak and has been observed experimentally in mixed-valent 
compounds such as CeOs$_4$Sb$_{12}$ and YbAl$_3$, is of intrinsic origin.  Detailed comparison is made between the resultant theory and transport/optical experiments on  
the filled-skutterudite compound CeOs$_4$Sb$_{12}$, and good agreement is obtained.

\end{abstract}



\section{Introduction.}
\label{sec:intro}

Kondo insulator compounds constitute a sub-class of lanthanide/actinide based heavy Fermion materials. Characterised by small spectral and optical gaps, and an activated resistivity at low temperatures, they have been under intense theoretical and experimental investigation
for many years (for reviews see \cite{grew91,hews, aepp,fisk,taka,degi,rise}).
From a theoretical perspective, the essential framework for understanding
heavy Fermion materials is the periodic Anderson model (PAM), wherein a single correlated 
$f$-level in each unit cell of the lattice hybridizes locally with a non-interacting
conduction band.

      Theories for Kondo insulators (KI) have generally been based on the 
particle-hole symmetric limit of the 
PAM~\cite{grew91,rise,prus95,geor,cmvarma,jarr,smit,vidh03}, in which 
the $f$-level occupancy ($n_f$) and the conduction band filling ($n_c$) are each
equal to unity. Although $n_f=1=n_c$ indeed satisfies the condition for an insulating ground state, it is not the generic case, which by contrast is  $n_f+n_c=2$ (as shown and discussed in~\cite{aepp,cmvarma,vidh04}). The strong coupling limit of the PAM is of course the Kondo lattice model with wholly localised $f$-electrons and hence $n_f =1$ necessarily, so one naturally expects that if a particular KI is sufficiently strongly correlated then the particle-hole symmetric PAM should provide a sound description of it. And indeed theories baseed on the symmetric
PAM have been able to describe, even quantitatively, many aspects of a number of these 
materials (see e.g.\ \cite{vidh03}).

  One cannot however expect such theories to be complete, since the fact that 
the generic condition for Kondo insulating behaviour is 
$n_f=2-n_c$~\cite{aepp,cmvarma,vidh04} itself suggests that 
KI materials are likely as a rule to be mixed-valent ($n_f\neq 1$); as indeed 
seems to be the case experimentally~\cite{aepp,rise}, and which behaviour lies beyond
the scope of the symmetric PAM. A more general treatment of KIs to encompass mixed-valent behaviour is thus clearly desirable, based on the general asymmetric PAM with the constraint $n_f=2-n_c\neq 1$. It is this we consider  here.

In a previous paper~\cite{vidh05-1}, working within the general framework
of dynamical mean-field theory (DMFT)~\cite{prus95,geor,voll,gebh},
we developed a local moment approach (LMA) (see~\cite{smit,vidh03,loga98,
glos,dick01,loga01,loga02,loga00,gloss03,bull} for details) to heavy Fermion systems. This
encompassed both strong coupling Kondo lattice behaviour ($n_f\rightarrow 1,n_c$ arbitrary), as well as the mixed-valence regime ($n_f$ and $n_c$ arbitrary). The theory was subsequently compared to experiments on several heavy Fermion metals, and excellent quantitative 
agreement was found~\cite{vidh05-2}. Although comparison was made only to metallic systems, 
the insulating ground state is just a particular solution of the same basic theory, that may be obtained by restricting the parameter set to the line $n_f+n_c=2$. In the present paper we treat the generic case by supplementing the theory of~\cite{vidh05-1} with a constraint on the total filling, $n_f+n_c=2$, but allowing $n_f$ and $n_c$ to deviate from unity.
This enables us to capture the mixed-valency along with the insulating nature of these systems.

We begin with a brief overview of the model and formalism used. 
Theoretical results for spectra and transport are discussed in section 3,
followed by a detailed comparison with experiments on the filled skutterudite
compound $CeOs_4Sb_{12}$, and a brief conclusion in section 4.

\section{Model and theory.}
\label{sec:model}

  In standard notation, the Hamiltonian for the PAM is given by:

\beqa
\hspace{-1cm}
\hat{H}=-t\sum_{(i,j),\sigma} c^\dag_{i\sigma} c^{\phantom{\dag}}_{j\sigma} + \ep_c\sum_{i,\sigma} c^\dag_{i\sigma} c^{\phantom{\dag}}_{i\sigma} \nnu \\ 
~~~ + \sum_{i,\sigma} (\ep_f + 
\case{U}{2} f^\dag_{i\,-\sigma}f^{\phantom{\dagger}}_{i\,-\sigma})
f^\dag_{i\sigma}f^{\phantom{\dag}}_{i\sigma}  \nnu \\
~~~~~~ + 
 V\sum_{i,\sigma} (f^\dag_{i\sigma} c^{\phantom{\dag}}_{i\sigma} +\mbox{h.c.})
\label{eq:model}
\eeqa
The first pair of terms describe the non-interacting conduction ($c$) band. The first gives the kinetic energy (or `free' conduction band $\hat{H}_{c}^{0}$), with the nearest neighbour hopping $t_{ij}=-t$ scaled as $t_{ij}\propto t^*/\sqrt{Z_c}$ 
in the large dimensional limit of coordination number $Z_c\rightarrow \infty$
~\cite{prus95,geor,voll,gebh}, and $t^*\equiv 1$ taken throughout as the unit
of energy; while the second gives the $c$-orbital energy, such that varying 
$\epsilon_{c}$ simply shifts the centre of gravity of the free conduction band relative to the Fermi level, and as such controls the conduction electron filling $n_{c}$.
The third term in $\hat{H}$ represents the $f$-orbital energy ($\epsilon_{f}$) and the on-site
Coulomb repulsion ($U$) for the localised $f$-orbitals, and the final term 
denotes the local hybridization between the $c$ and $f$-electrons which is 
responsible for making the otherwise localized $f$-electrons itinerant.

  The free conduction band may be diagonalised,
$\hat{H}_{c}^{0} \equiv \sum_{\mathbf{k},\sigma}\epsilon^{\phantom\dagger}_{\mathbf{k}}c^{\dagger}_{\mathbf{k}\sigma}c^{\phantom\dagger}_{\mathbf{k}\sigma}$,
with corresponding density of states $\rho_{0}(\epsilon)= N^{-1}\sum_{\mathbf{k}}\delta (\epsilon -\epsilon_{\mathbf{k}})$.
While the basic formalism is valid for any underlying lattice
and associated $\rho_0(\ep)$, we consider in this paper
the specific case of the hypercubic lattice 
(which is Bloch-decomposable, unlike e.g.\ a Bethe lattice), for which
$\rho_0(\ep)=\exp(-\ep^2)/\sqrt{\pi}$ is a Gaussian. \\

Within DMFT, the $f$-electron self-energy $\Sigma_f(\om;T)$, 
representing many-body scattering due to electron interactions,
is rendered purely local (i.e.\ site-diagonal or momentum independent).
The local, site-diagonal Green functions  for the conduction-
($G^c(\om)$) and $f$-electrons ($G^f(\om)$)  
are then given by~\cite{prus95,geor,voll,gebh,vidh04}

\numparts
\beqa
\fl
G^c(\om)&=& \int^{\infty}_{-\infty}d\epsilon ~ \frac{\rho_{0}(\epsilon)}{\om^+ -\epsilon_{c} - \frac{V^{2}}{\om^+ - \epsilon_{f} - \Sigma_{f}(\om;T)} - \epsilon}~~\equiv ~~ \int^{\infty}_{-\infty}d\epsilon ~ \rho_{0}(\epsilon)~G^{c}(\epsilon;\om)
\label{eq:gc} \\
\fl
G^{f}(\om)&=& \frac{1}{\om^+ - \epsilon_{f} - \Sigma_{f}(\om;T)} \left [ 1 +
\frac{V^{2}}{\om^+ - \epsilon_{f} - \Sigma_{f}(\om;T)}~G^c(\om;T) \right]
\label{eq:gf}
\eeqa
\endnumparts 
(with $\om^+ = \om + i0^+$). Solution of these DMFT equations naturally requires a knowledge -- and self-consistent determination -- of the $f$-electron self-energy $\Sigma_f(\om;T)$, which DMFT by itself does not of course prescribe. For this, we employ the physically transparent local moment approach (LMA)~\cite{loga98,glos,dick01,loga01,loga02,loga00,gloss03,bull,vidh03,vidh05-2,smit,vidh05-1,vidh04}. This handles non-perturbatively all interaction strengths from weak to strong coupling, and all relevant energy/temperature scales, while at the same time recovering the dictates of Fermi liquid behaviour at low energy/temperature scales. Originally introduced to describe Anderson impurity models~\cite{loga98,glos,dick01,loga01,loga02,loga00,gloss03,bull}, for which it is found to agree well with e.g.\ numerical renormalization group calculations and a number of exact results,
lattice-based heavy Fermion systems have also been considered within DMFT+LMA~\cite{vidh03,vidh05-1,smit,vidh05-2,vidh04}. 

Here we sketch in brief only the essential elements of the LMA:

\noindent
(i)The starting point is simple \emph{static} mean-field, 
i.e.\ unrestricted Hartree-Fock. 
This has the virtue of recognising local moment formation as the first effect of electron interactions, by introducing the possibility of local moments from the outset. But by itself it is inadequate, for two reasons. First, it results in a (locally doubly-degenerate) symmetry broken mean-field state, which is not perturbatively connected (in $U$) to the 
non-interacting limit and in consequence violates Fermi liquid behaviour at low energies. Second, its inherently static nature cannot by construction capture electron correlation effects. It is these signal limitations the LMA overcomes.

\noindent
(ii) 
Electron correlations, embodied in dynamical self-energies, are incorporated
within the framework of a spin-rotationally invariant \emph{two-self-energy} description which is an inevitable consequence of the underlying two mean-field saddle points (and from which the conventional single self-energy $\Sigma_{f}(\om;T)$ is recovered simply as a byproduct). The resultant dynamical self-energies are built diagrammatically from, and are self-consistently determined functionals of, the underlying mean-field propagators. They include in particular a non-perturbative class of diagrams that capture the spin-flip dynamics central to the physics of the PAM.

\noindent
(iii)The third, key element of the LMA is that of \emph{symmetry 
restoration}~\cite{smit,glos,dick01,gloss03}: self-consistent restoration of the symmetry broken at pure mean-field level, and hence correct recovery of the low-energy local Fermi liquid behaviour that reflects adiabatic continuity to the non-interacting limit.
This is embodied in a single self-consistency condition on the two-self-energies precisely at the Fermi level ($\om =0$), which in practice amounts to a self-consistent determination of the local moment (supplanting the simple `gap equation' for such that arises at crude mean-field level).

\noindent
(iv) Luttinger's theorem~\cite{hews,lutt}, itself a reflection of perturbative continuity to the non-interacting limit
(i.e.\
$I_{L} = \mathrm{Im}\int^{0}_{-\infty}d\omega ~(\partial\Sigma_{f}(\om;T=0)/\partial\om)G^{f}(\omega) ~=~0$), is also satisfied by the LMA~\cite{vidh04}. We note too that adiabatic continuity to the non-interacting limit is characteristic of 
\emph{both} the heavy Fermion (metallic) and Kondo insulating states~\cite{smit,vidh03}.

  Full details regarding the structure and implementation of the Local Moment Approach for the generic asymmetric PAM are given in~\cite{vidh05-1,vidh04}, to which the reader is referred for further information.\\

    Since our objective is to study generic Kondo insulators, corresponding
to $n_{c}+n_{f} =2$ (for $T=0$ where the distinction between an insulator and 
a metal has strict meaning), we now examine the general conditions under 
which an insulating ground state is obtained.  To this end we note that
the Luttinger theorem $I_{L}=0$ can be expressed as the following exact 
statement for the Fermi surface of the PAM, as shown in~\cite{vidh04}:
\beq
\case{1}{2}(n_c+n_f) = \int^{-\ep_c + 1/\tilde{\ep}_f^*}_{-\infty} \rho_0(\ep)
\,d\ep\; + \;\theta(-\tilde{\ep}_f^*)
\label{eq:lutt}
\eeq
(with $\theta(x)$ the unit step function). Here $\tilde{\ep}_{f}^*=\ep_f^*/V^2$,
and $\epsilon_{f}^*$ is the renormalised or effective $f$-level given by
$\ep_f^*=\ep_f + \Sigma_f^R(\om=0;T=0)$ (where $\Sigma_f^R = \mathrm{Re}\Sigma_{f}$).

From this it follows that, for any conduction band $\rho_{0}(\epsilon)$, the KI filling constraint $n_{c}+n_{f} = 2$ is satisfied if $\epsilon_{f}^* =0$ (whether
$\epsilon_{f}^* = 0^+$ or $0^-$). Moreover for a non-compact bare density of states
$\rho_{0}(\epsilon)$, such as the Gaussian appropriate to the hypercubic lattice we consider, 
$\epsilon_{f}^* = 0$ is the only possibility that satisfies $n_{c}+n_{f} =2$.
As such, $\epsilon_{f}^* =0$ is the general condition for a KI that we seek;
with the (particle-hole) symmetric KI ($\epsilon_{f} = -U/2$ and
$\epsilon_{c} =0$) simply the particular case for which $n_{c}=1=n_{f}$.

  The condition $\epsilon_{f}^* =0$ also leads naturally 
to a gap at the Fermi level ($\om =0$) in the $T=0$ single-particle spectra 
$D^{c}(\om)$ and $D^{f}(\om)$. This is most easily seen from the limiting low-frequency behaviour of equations 
(\ref{eq:gc}) and (\ref{eq:gf}) for $T=0$, obtained from a low-$\om$ quasiparticle
expansion~\cite{vidh04} and given by
\numparts
\beqa
D^c(\om)\sim\rho_0(-\ep_c-\frac{1}{\omt - {\tilde{\ep}_{f}}^*}) \label{ad1}\\
V^2D^f(\om)\sim\frac{1}{(\omt-\tilde{\ep}_{f}^{*})^2}\rho_0(-\ep_c-\frac{1}{\omt- {\tilde{\ep}_{f}}^*}) \,
\label{ad2}
\eeqa
\endnumparts
where $\omt=\om/\om_L$. Here, $\om_L=ZV^2/t_*$ is
the characteristic low-energy Fermi liquid scale in the problem (with
$Z =[1-(\partial\Sigma^{R}_{f}(\om; T=0)/\partial\om)_{\omega =0}]^{-1}$ the
usual quasiparticle weight, or inverse mass renormalization factor).
From this it is readily seen that with $\epsilon_{f}^{*} =0$, $D^{c}(\om =0) =0$ (and likewise for
the $f$-spectrum), as one expects for an insulator.

  In this paper we implement the filling condition $n_{c}+n_{f}=2$ simply
by supplementing the LMA with the $T=0$ constraint 
$\ep_f^*=\ep_f+\Sigma_f^R(\om=0;T=0)=0$
(which is algorithmically simple), ensuring thereby direct access to the generic Kondo insulating states of interest. The LMA yields directly the local Green functions ($G^c$ and $G^{f}$) and self-energies, knowledge of which is well known to be sufficient within DMFT~\cite{prus95,geor,voll,gebh} to determine d.c.\ transport and optical properties, as detailed e.g.\ in~\cite{vidh03,vidh05-1}.
In the next section, we discuss our theoretical results.

\section{Theoretical results and discussion}

The PAM, equation~\ref{eq:model}, is characterised by four `bare' material
parameters, $U, V, \epsilon_c$ and the $f$-level asymmetry  
$\eta = 1+2\epsilon_{f}/U$ (or equivalently
$\epsilon_{f}$ itself). The local spectra naturally depend on these parameters,
and the ($T=0$) total filling is given by
\beq
\hspace{-1cm}
n_c+n_f~=~2\int^0_{-\infty}d\om\,[D^{c}(\om;U,V,\ep_c,\eta)
+D^{f}(\om;U,V,\ep_c,\eta)]\,.
\label{eq:fconst}
\eeq
For a metallic phase, 
$(U,V,\epsilon_{c},\eta)$ are in general independent parameters. But for the 
Kondo insulating state, the constraint
$n_{c}+n_{f} =2$ obviously implies that only 3 of the bare parameters are 
independent, e.g.\ $\eta \equiv \eta(U,V,\epsilon_{c})$.

\begin{figure}[t]
\begin{center}
\includegraphics[height=6cm,clip]{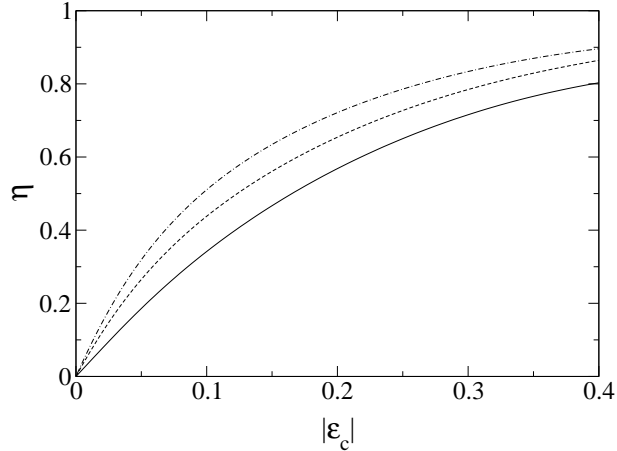}
\end{center}
\caption{Variation of $f$-level asymmetry $\eta$ with $|\epsilon_{c}|$ 
($=-\epsilon_{c}$) for the Kondo insulating state, for fixed $V^{2} =0.2$ and three different interactions $U= 2.6$ (solid line), $3.8$ (dashed line) and $5.2$ (point-dash). With increasing $|\epsilon_{c}|$, the insulator becomes progressively more
mixed-valent (the $f$-electron filling dropping from $n_{f} =1$ for 
$|\epsilon_{c}| =0$ to $n_{f} \simeq 0.63$ for $|\epsilon_{c}|=0.4$).
}
\label{fig:Uv2}
\end{figure}

We illustrate this is figure~\ref{fig:Uv2}, showing the resultant $\eta$ 
\emph{vs} $|\epsilon_{c}|$ (with $\epsilon_{c} \leq 0$ and hence $n_{c} \ge 1$), for a fixed value of the hybridization $V$ and three different interactions $U$.
All curves meet at the origin, this being the symmetric KI ($n_{f}=1=n_{c}$) for which $\eta =0 =\epsilon_{c}$ for \emph{all} $U$ and $V$, i.e.\ $\eta(U,V,\epsilon_{c}=0) =0$ is independent of $U$ and $V$. Away from $\epsilon_{c} =0$ however, $\eta$ depends generically on all of $U, V$ and $\epsilon_{c}$. On increasing $|\epsilon_{c}|$ for any given $U$ and 
$V$, as in figure~\ref{fig:Uv2}, the system becomes progressively mixed-valent:
$\eta$ increases (i.e.\ the $f$-level moves upwards towards the Fermi level),
and $n_{f}$ steadily diminishes from $n_{f}=1$ at $\epsilon_{c}=0$, with a concomitant increase in the resultant quasiparticle weight/inverse mass renormalization $Z$ (and hence in the low-energy scale $\om_{L} = ZV^2$). In fact over the $U$-range shown in figure~\ref{fig:Uv2}, the resultant $n_{f}$s for given $\epsilon_{c}$ barely change with $U$, $n_{f}$ dropping to $\simeq 0.63$ for $|\epsilon_{c}| =0.4$.

\begin{figure}[h]
\epsfxsize=300pt
\begin{center}
\epsffile{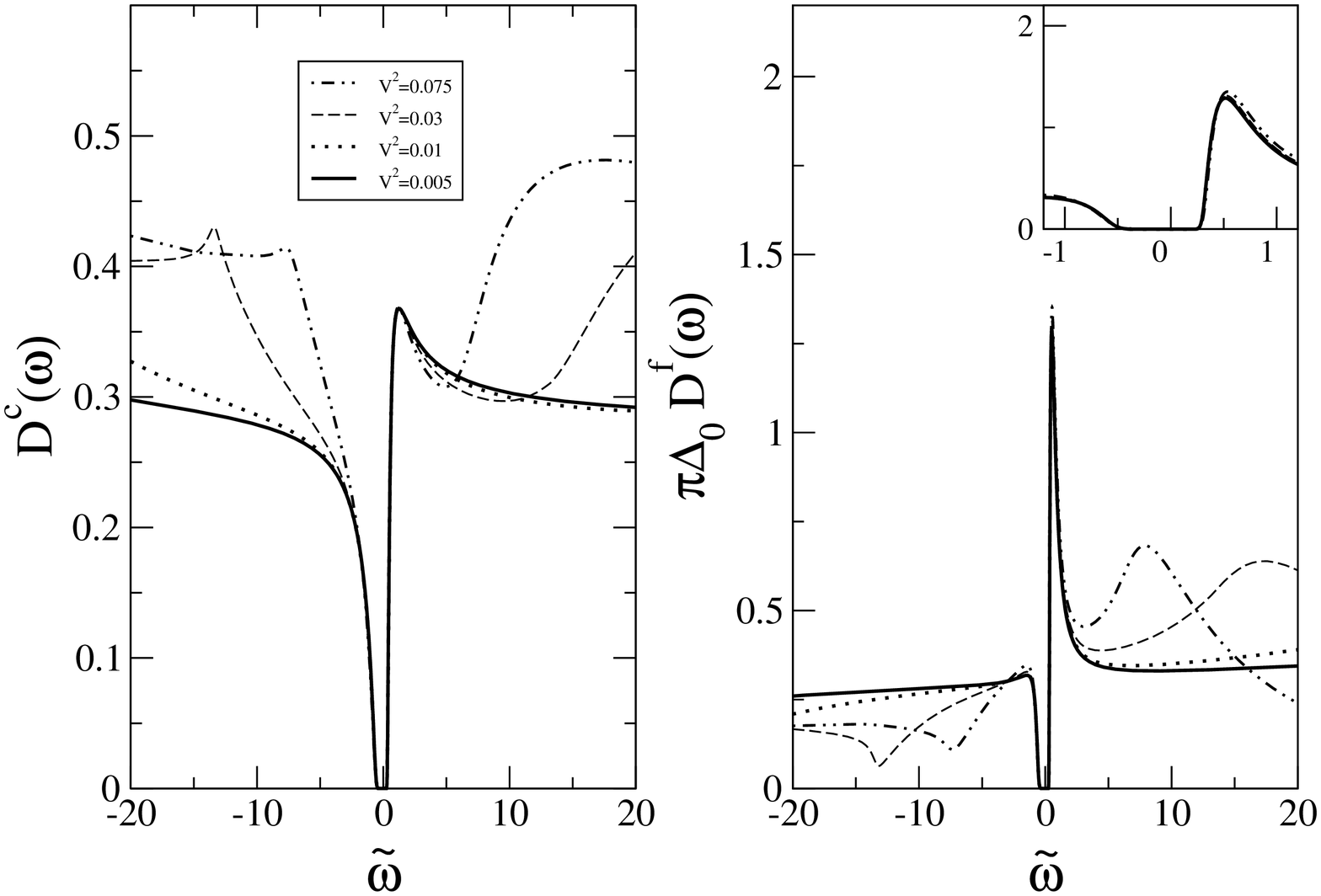}
\end{center}
\caption{
The local conduction electron spectra (left panel) and $f$-electron
spectra (right panel, with the constant $\Delta_{0}=\pi V^{2}\rho_{0}(-\epsilon_{c})$) are shown {\it vs.\ }$\omt=\om/\om_L$ 
for a fixed $U=2.0, \ep_c=-0.4$ and varying
$V^2=0.005$ (solid line), 0.01 (dotted), 0.03 (dashed) 0.075 (double dot dashed).
An asymmetric gap is seen straddling the Fermi level. Inset: 
magnified view of the low frequency $f$-spectrum.}
\label{fig:spec}
\end{figure}

To illustrate typical local single-particle dynamics, figure~\ref{fig:spec} shows
the $T=0$ LMA $c$-and $f$-spectra \emph{vs} $\omt=\om /\om_L$ (left and right panels 
respectively) for a fixed $\ep_c=-0.4$, $U=2$ and a range of different hybridization couplings
$V$ as indicated. The associated $n_{f}$ differs only slightly with $V$ over the range shown, with $n_{f} \simeq 0.6$ indicating mixed-valent character. The single-particle spectra for this representative asymmetric KI are indeed seen to be gapped at the Fermi
level, with magnitude $\sim \om_{L}$, and the obvious key point is that the gap is asymmetric about the Fermi level (it is of course strictly soft for a hypercubic lattice, but with exponentially small spectral density in the vicinity of the Fermi level). 
The four sets of spectra are seen to be quite distinct away from the Fermi level, as one expects. However the limiting low-frequency spectral forms are given (asymptotically exactly) by 
equations~(\ref{ad1},~\ref{ad2}) which, with $\epsilon_{f}^* =0$ as appropriate to  the KI, are seen to depend solely on $\epsilon_{c}$ and $\tilde{\om}$; i.e.\ they should be independent of $V$ or $U$ (which we note has nothing \emph{per se} to do with the `universal scaling' of spectra characteristic of the strong coupling limit
$n_{f} \simeq 1$~\cite{smit,vidh03,vidh04,vidh05-1}). That the LMA correctly recovers this behaviour correctly is self-evident from the figure.

\subsection{Optical conductivity.}

\begin{figure}[t]
\begin{center}
\includegraphics[height=6cm,clip]{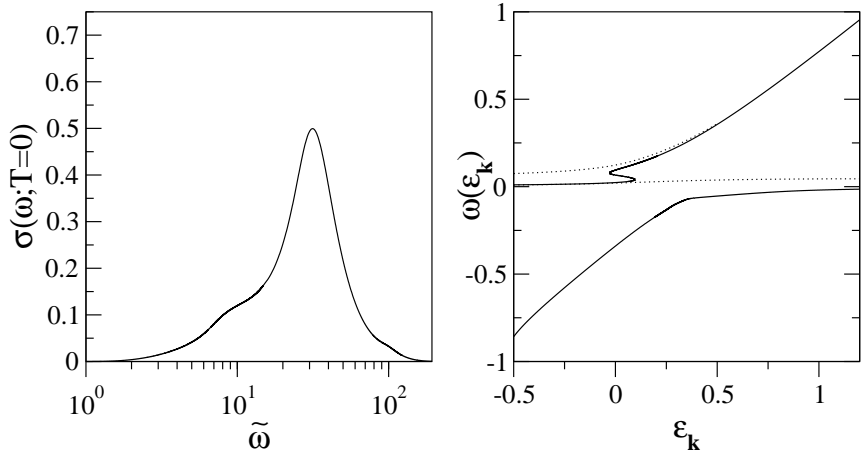}
\end{center}
\caption{{\emph{Left panel}: Optical conductivity $\sigma(\om;0)$ \emph{vs} $\tilde{\om} = \om/\om_{L}$ for a representative mixed-valent Kondo insulator ($n_{f} \simeq 0.73$), with
$U=3.8, V^2 =0.2, \epsilon_{c} = -0.3$ and $\eta = 0.8$.
\emph{Right panel}: the renormalized bandstructure $\omega(\epsilon_{\mathbf{k}})$ \emph{vs} $\epsilon_{\mathbf{k}}$ (solid lines). For the dotted lines, see discussion in text.
}}
\label{fig:fig3}
\end{figure}

  We consider now the frequency dependence of the optical conductivity 
$\sigma (\om;T)$, focusing specifically on $T=0$ (results at 
finite-$T$ will be included in the following section). Figure~\ref{fig:fig3} shows the optical conductivity obtained from the LMA for a representative mixed-valent Kondo insulator. The dominant feature in the optics is the usual strong, direct gap absorption (`mid-infrared peak'). In the absence of scattering due to electron interactions there would be no absorption whatever below the direct gap~\cite{vidh03,vidh05-1}; but, just as for the (particle-hole) symmetric KI considered in~\cite{vidh03}, interaction-induced many-body scattering leads as seen to continuous absorption all the way down to the indirect gap scale 
$\om = \Delta_{\mathrm{ind}} = 2\om_{L}$ (which is why $\sigma (\om;0)$ is 
shown as a function of $\tilde{\om} =\om/\om_{L}$, figure~\ref{fig:fig3}
showing there is indeed negligible absorption below $\tilde{\om} \simeq 2$).

  Figure~\ref{fig:fig3} also shows that the optical conductivity possesses a distinct low-frequency shoulder, lying somewhat above the indirect gap (at 
$\tilde{\om} \approx 10$ in the present example). Such a feature occurs neither in the non-interacting limit nor for the interacting symmetric KI~\cite{vidh03}. From investigation of a wide range of underlying material parameters however, we find it to be entirely typical of interacting mixed-valent (necessarily asymmetric) Kondo insulators, albeit naturally becoming less pronounced as the symmetric limit is approached.
In previous work~\cite{vidh05-1,vidh05-2} we have also found such behaviour to be characteristic of correlated intermediate valence metals. We conclude therefore that a low-frequency shoulder should typically exist,
as an \emph{intrinsic} optical feature, in interacting mixed-valent systems
whether metallic or insulating; and that it arises from a combination of many-body scattering and mixed-valency.
Experimentally, such a shoulder has been observed in the optical conductivity of  Kondo insulators such as CeOs$_{4}$Sb$_{12}$~\cite{cosopt} as well as in intermediate valence metals such as YbAl$_{3}$~\cite{ybal}.
In~\cite{vidh05-2} we considered the case of YbAl$_{3}$ in detail, and obtained very good agreement between theory and experiment, including striking reproduction of the low-energy optical shoulder. Analogous comparison for CeOs$_{4}$Sb$_{12}$ will be given in the following section.

  First, however, we seek to interrogate the optics in more microscopic detail. In the absence of vertex corrections (as appropriate to DMFT~\cite{prus95,geor,voll,gebh}) the $T=0$ optical conductivity is given by~\cite{vidh03,vidh05-1}
\numparts
\beq
\hspace{-1cm}
\sigma (\om; 0) \propto \int^{\infty}_{-\infty} d\epsilon~\rho_{0}(\epsilon)~I(\epsilon;\om)
\label{eq:optcon1}
\eeq
(bar extraneous constants), with 
\beq
\hspace{-1cm}
I(\epsilon;\om) ~=~ \frac{1}{\om} \int^{0}_{-\om}d\om_{1}~D^{c}(\epsilon;\om_{1})D^{c}(\epsilon;\om_{1}+\om) \,.
\label{eq:optcon2}
\eeq
\endnumparts
Here, $D^{c}(\epsilon;\om) \equiv D^{c}(\epsilon_{\mathbf{k}}=\epsilon;\om)$ is the 
$\epsilon_{\mathbf{k}}$-resolved conduction electron spectrum (accessible e.g.\ via ARPES): $D^{c}(\epsilon;\om) = -(1/\pi)\mathrm{Im}G^{c}(\epsilon;\om)$,
with the $\mathbf{k}$-space propagator $G^{c}(\epsilon_{\mathbf{k}};\om)$ given by
\numparts
\beq
\hspace{-1cm}
G^{c}(\epsilon_{\mathbf{k}};\om) = [\gamma(\om) - \epsilon_{\mathbf{k}}]^{-1}
\label{eq:optcon3}
\eeq
where
\beq
\hspace{-1cm}
\gamma(\om) = \omega^{+} - \epsilon_{c} - \frac{V^{2}}{\om^{+}-\epsilon_{f} -\Sigma_{f}(\om;T)}
\label{eq:optcon4}
\eeq
\endnumparts
(such that the local, site-diagonal propagator $G^{c}(\om) = N^{-1}\sum_{\mathbf{k}}G^{c}(\epsilon_{\mathbf{k}};\om) \equiv \int d\epsilon ~ \rho_{0}(\epsilon)G^{c}(\epsilon;\om)$ as in equation ~(\ref{eq:gc})).

  The physical content of equations (3.2) is clear: for given $\epsilon \equiv \epsilon_{\mathbf{k}}$ (as in equation(~\ref{eq:optcon2})), optical transitions
occur from a state below the Fermi level
(lying at $\om_{1} < 0$) to a state $\om$ higher in energy that lies 
above the Fermi level (at $\om_{1}+\om >0$); all transitions being `direct' in
the sense that absorption occurs for given/fixed $\epsilon_{\mathbf{k}}$. The net
optical conductivity at frequency $\om$ is then, as in equation~(\ref{eq:optcon1}),
the sum of all such processes over the full range of $\epsilon \equiv \epsilon_{\mathbf{k}}$ (with density $\rho_{0}(\epsilon) = N^{-1}\sum_{\mathbf{{k}}} \delta (\epsilon_{\mathbf{k}} - \epsilon)$).
This picture is of course wholly familiar at an elementary level in 
the context of non-interacting electrons (or `renormalized' non-interacting electrons, as considered below).
But we emphasise that it is quite general:
it applies equally well to the fully interacting case, where the states between which absorptive transitions occur are many-body states. There is of course an obvious difference between the interacting and non-interacting cases, namely the existence of scattering (and hence `lifetime' effects) due to electron interactions in the former case. That in turn generates a profound difference between optics in the two cases, which we now consider since
it also throws light on the optical conductivity shown in figure~\ref{fig:fig3}.

  The simplest description of the optics is at the level of `renormalized bandstructure'~\cite{geor}. Here, $\gamma(\om)$ in equation~(\ref{eq:optcon3}) is taken to be purely real,
$\gamma(\om) \equiv \gamma^{R}(\om)$, neglect of $\gamma^{I}(\om) = \mathrm{Im}\gamma(\om)$ meaning from equation~(\ref{eq:optcon4}) that the imaginary part of the $f$-electron self-energy --- the source of all scattering --- is completely neglected.
In addition, $\gamma^{R}(\om)$ is further approximated by its asymptotic 
low-frequency behaviour, which (from equation~(\ref{eq:optcon4})) comes from that for
$\Sigma_{f}^{R}(\om;0) \equiv \mathrm{Re}\Sigma_{f}(\om;0)$, namely the
simple Taylor expansion
$\Sigma_{f}^{R}(\om;0) \sim  \Sigma_{f}^{R}(0;0) -(1/Z -1)~\om$ (with $Z$ the usual quasiparticle weight). With this approximation
$G^{c}(\epsilon_{\mathbf{k}};\om) \simeq [\gamma^{R}(\om) - \epsilon_{\mathbf{k}}]^{-1}$ has poles at two frequencies,
$\om = \om^{+}(\epsilon_{\mathbf{k}}) ~>0$ above the Fermi level and
$\om = \om^{-}(\epsilon_{\mathbf{k}})~<0$ below it. These are given explicitly by
\numparts
\beq
\om^{\pm}(\epsilon_\k)=\case{1}{2}\left[(\ep_c+\ep_\k)\pm\sqrt{(\ep_c+\ep_\k)^2
+4ZV^2}\right]
\label{eq:qp1}
\eeq
(where we use the fact that the renormalized level $\epsilon_{f}^{*} = \epsilon_{f}+\Sigma^{R}_{f}(0;0) ~=0$ for a generic KI, as discussed in section 2), with a gap between them $\tilde{\Delta}(\epsilon_{\mathbf{k}}) = \om^{+}(\epsilon_\k) -
\om^{-}(\epsilon_\k)$ of
\beq
\tilde{\Delta}(\epsilon_{\mathbf{k}}) = \sqrt{(\ep_c+\ep_\k)^2 +4ZV^2}\,.
\label{eq:qp2}
\eeq
\endnumparts
Equation~(\ref{eq:qp1})) is simply the renormalized bandstructure of the problem (reducing trivially to the non-interacting limit result for $Z=1$), the two branches
$\om^{+}(\epsilon_{\mathbf{k}})$ and $\om^{-}(\epsilon_{\mathbf{k}})$ reflecting 
physically the fact that, per unit cell, a single $f$-level hybridizes locally to a single conduction band. The minimum gap between them, the `direct gap' at this level, occurs for
$\epsilon_{\mathbf{k}} = -\epsilon_{c}$ and is thus $\Delta_{\mathrm{dir}} =
 \tilde{\Delta}(\epsilon_{\mathbf{k}}=-\epsilon_{c}) ~=~ 2\sqrt{Z}V$.

\begin{figure}[t]
\begin{center}
\includegraphics[height=6cm,clip]{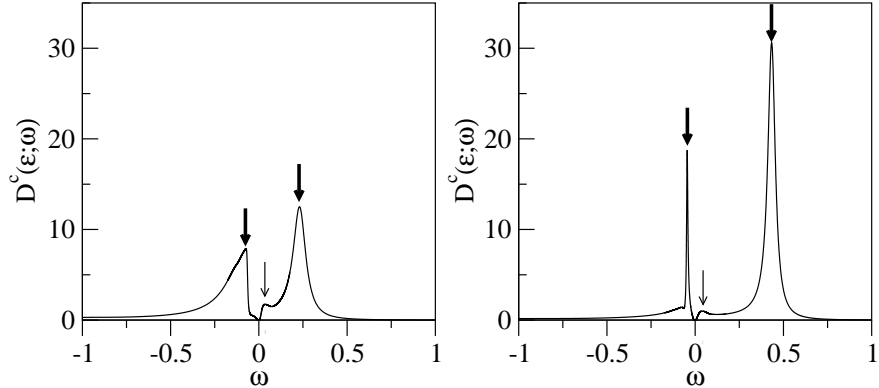}
\end{center}
\caption{{For the same bare parameters as in figure~\ref{fig:fig3}.
\emph{Left panel}: $T=0$ conduction electron spectrum 
$D^{c}(\epsilon_{\mathbf{k}};\om)$
\emph{vs} $\om$, for fixed $\epsilon_{\mathbf{k}} = 0.3$ ($=-\epsilon_{c}$).
The solid arrows indicate the location of the pure poles that would arise 
from a simple renormalized band picture (following from the renormalized  
bandstructure shown in figure~\ref{fig:fig3}, right panel). An additional spectral feature above the Fermi level, arising from interactions, is indicated by a single arrow. \emph{Right panel}: The same, shown for an $\epsilon_{\mathbf{k}} = 0.6$.
}}
\label{fig:fig4}
\end{figure}

Since the resultant $D^{c}(\epsilon_{\mathbf{k}};\om') \propto \mathrm{Im}G^{c}(\epsilon_{\mathbf{k}};\om')$ contains two poles, at $\om' = \om^{+}(\epsilon_{\mathbf{k}})$ and $\om^{-}(\epsilon_{\mathbf{k}})$, it follows from equation~(\ref{eq:optcon2}) that 
optical absorption can only arise from transitions between these two $\delta$-peaks
--- provided of course their separation $\om^{+}(\epsilon_\k) -\om^{-}(\epsilon_\k)$ concides with the requisite absorption frequency $\om$; and in consequence that there is no absorption at all (i.e.\ $\sigma(\om;0) =0$)
for $\om \leq \Delta_{\mathrm{dir}} =
2\sqrt{Z}V$. This level of description may be improved, very slightly, by
retaining
$G^{c}(\epsilon_{\mathbf{k}};\om) \simeq [\gamma^{R}(\om) - \epsilon_{\mathbf{k}}]^{-1}$ but with the full LMA $\gamma^{R}(\om)$ employed instead of its asymptotic low-$\om$ expansion. The results of such a calculation are shown in figure~\ref{fig:fig3} (right panel, solid lines, where the `band branching' seen over a narrow range around $\epsilon_{\mathbf{k}}=0$ is simply a consequence of retaining $\gamma^{R}(\om)$ but neglecting $\gamma^{I}(\om)$).

  The above `renormalized band' description, commonly employed though it is, is qualitatively inadequate. The picture changes markedly when interaction-induced scattering is properly
retained, as embodied in the imaginary part of the $f$-electron self-energy and hence 
a non-vanishing conduction electron scattering rate $\gamma^{I}(\om)$. 
To illustrate this, figure~\ref{fig:fig4} shows the full LMA conduction electron spectra
$D^{c}(\epsilon_{\mathbf{k}} ;\om)$ \emph{vs} $\om$ for two values of $\epsilon \equiv \epsilon_{\mathbf{k}}$, viz $\epsilon_{\mathbf{k}}= -\epsilon_{c} = 0.3$ 
(left panel) and $\epsilon_{\mathbf{k}} =0.6$ (right panel). At the simplistic level of renormalized bands, as in the right panel to figure~\ref{fig:fig3}, each of these spectra would consist of one pole below the Fermi level ($\om =0$) and one above it.
Reality is clearly different: the spectra are significantly broadened due to 
interactions and form a continuum (save for the expected gap in the immediate vicinity of the Fermi level).
It is for this reason that, in contrast to the renormalized band picture, conduction electron spectra for essentially \emph{any} $\epsilon_{\mathbf{k}}$ contribute to optical absorption for \emph{all} frequencies down to the lowest energy (indirect) gap scale.

  In figure~\ref{fig:fig4} we also mark (by solid arrows) the locations in $D^{c}(\epsilon_{\mathbf{k}};\om)$ of the two nominal poles that would arise at the renormalized band level (their positions can be read off from 
the right panel in figure~\ref{fig:fig3} at the appropriate $\epsilon_{\mathbf{k}}$). These are seen to correspond
rather accurately to the position of the dominant peak maxima in $D^{c}(\epsilon_{\mathbf{k}};\om)$. Just above the Fermi level, an additional small peak
(marked by an arrow) can also be seen in figure~\ref{fig:fig4} . This feature does not of course have any counterpart at the renormalized band level, and we find its existence to be characteristic of mixed-valent Kondo insulators (it does not occur in the particle-hole symmetric limit). These spectral features, two peaks above and one below the Fermi level, are found to be characteristic of the conduction spectra
$D^{c}(\epsilon_{\mathbf{k}};\om)$ for all $\epsilon_{\mathbf{k}}$.
The positions of the two peak maxima above the Fermi level in the full
$D^{c}(\epsilon_{\mathbf{k}};\om)$ can be mapped out as a function of 
$\epsilon_{\mathbf{k}}$, and are shown in figure~\ref{fig:fig3} (right panel, dotted lines). Note in particular that the lower-energy such peak lies close to, but slightly above, the Fermi level for all $\epsilon_{\mathbf{k}}$. We thus expect
(see equation~(\ref{eq:optcon2})) significant optical absorption \emph{to} this peak \emph{from} the dominant spectral peak in 
$D^{c}(\epsilon_{\mathbf{k}};\om)$ below the Fermi level; particularly at 
relatively low frequencies where (see figure~\ref{fig:fig3} right panel) 
for $\epsilon_{\mathbf{k}} \gtrsim 0.3$ or so the latter peak itself lies close
to the Fermi level, exhibits only modest dispersion with $\epsilon_{\mathbf{k}}$, and
is spectrally sharp (as in figure~\ref{fig:fig4} right panel).

\begin{figure}[t]
\includegraphics[height=6cm,clip]{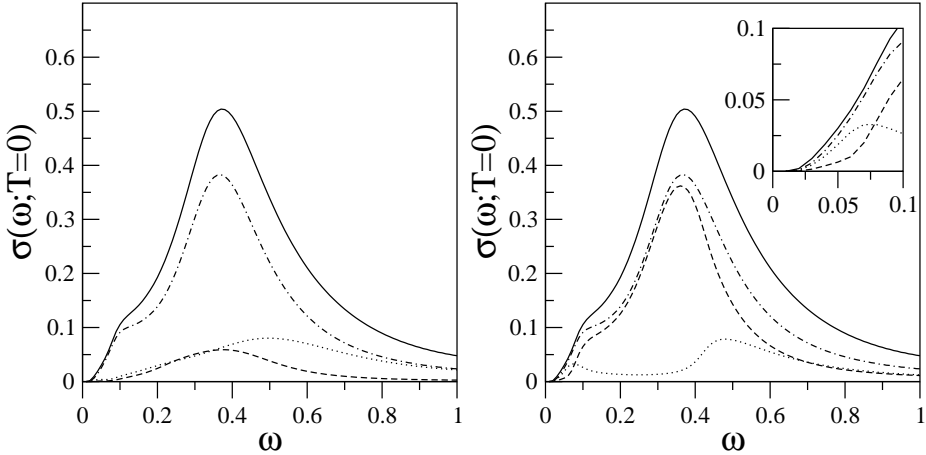}
\begin{center}
\end{center}
\caption{{For the same bare parameters as in figure~\ref{fig:fig3}.
\emph{Left panel}: the $T=0$ optical conductivity $\sigma(\om;0)$ \emph{vs} $\om$ (solid line) and the contribution to it arising from $\epsilon \equiv \epsilon_{\mathbf{k}} > 0.1$ (point-dash), $0< \epsilon_{\mathbf{k}} < 0.1$ (dashed) and $\epsilon_{\mathbf{k}} <0$ (dotted). \emph{Right panel}: showing further the contributions to
$\sigma(\om;0)$ from $0.1 < \epsilon_{\mathbf{k}} < 0.5$ (dashed line) and
$\epsilon_{\mathbf{k}} >0.5$ (dotted). Inset: shown on a lower-$\om$ scale.
 }}
\label{fig:fig5}
\end{figure}

We return now to the optical conductivity 
$\sigma(\om;0)$ given by equations~(\ref{eq:optcon1}, \ref{eq:optcon2}), shown in the left panel of figure~\ref{fig:fig3} for the representative bare parameters indicated, and
again on a linear-$\om$ scale in figure~\ref{fig:fig5} (solid line). With the 
preceding comments in mind, our aim is to determine
what range of values of $\epsilon \equiv \epsilon_{\mathbf{k}}$ give the primary contribution to $\sigma(\om;0)$ in different frequency intervals --- in particular, frequencies
in the vicinity of (a) the dominant direct gap absorption ($\om \approx 0.4$ in the present example), (b) the low-frequency shoulder ($\om \approx 0.1$), and (c) the lowest frequency scales down to the indirect gap ($\om \approx 0.02$).

  To that end, we simply partition the 
$\epsilon$-integral in equation~(\ref{eq:optcon1}) for $\sigma(\om;0)$, into contributions from different $\epsilon$-intervals. This is shown in the left panel to figure~\ref{fig:fig5}, where the separate contributions to the total 
$\sigma(\om;0)$ (solid line) from $\epsilon \equiv \epsilon_{\mathbf{k}} > 0.1$ (point-dash line), $0< \epsilon_{\mathbf{k}} < 0.1$ (dashed) and 
$\epsilon_{\mathbf{k}} <0$ (dotted) are shown.
From this it is clear that the dominant contribution to $\sigma(\om;0)$ in all three $\om$-regimes of interest arises from $\epsilon_{\mathbf{k}} > 0.1$ (the associated renormalized bandstructure being shown in the right panel of figure~\ref{fig:fig3}). To distinguish the $\om$-regimes, the right panel in figure~\ref{fig:fig5} shows the contribution to $\sigma(\om;0)$ from $0.1 < \epsilon_{\mathbf{k}} < 0.5$ (dashed line) and $\epsilon_{\mathbf{k}} >0.5$ (dotted). From the renormalized band  description discussed above, the direct gap occurs at $\epsilon_{\mathbf{k}} = -\epsilon_{c}$ ($=0.3$ in the present example); and consistent both with this and the inevitable broadening induced by scattering, the dominant contribution to $\sigma(\om;0)$ in the vicinity of the strong direct gap 
absorption at $\om \approx 0.4$ is indeed seen to arise from the interval $0.1 < \epsilon_{\mathbf{k}} < 0.5$ --- i.e.\ $\epsilon_{\mathbf{k}} = -\epsilon_{c} \pm 0.2$. In contrast, as seen in particular from the right inset to figure~\ref{fig:fig5}, absorption on the lowest frequencies down to the order of the indirect gap is controlled by $\epsilon_{\mathbf{k}}\gtrsim 0.5$. This in turn is consistent with the form of $D^{c}(\epsilon_{\mathbf{k}};\om)$ for $\epsilon_{\mathbf{k}}$'s in this range, exemplified by the right panel in 
figure~\ref{fig:fig4} and discussed above. Finally, the low-$\om$ shoulder 
in $\sigma(\om;0)$ (at $\om \approx 0.1$) that is typical of the mixed-valent Kondo insulator, is seen to stem mainly from the interval $0.1 < \epsilon_{\mathbf{k}} < 0.5$, its existence reflecting the small spectral feature in $D^{c}(\epsilon_{\mathbf{k}};\om)$ above the Fermi level discussed above, and shown in figure~\ref{fig:fig4}.

\section{Comparison to transport in CeOs$_4$Sb$_{12}$}

The filled-skutterudite compounds RT$_4$X$_{12}$ (R=rare earth, T=Transition
metal and X=pnictide) have attracted much experimental interest, due in part 
to their possible applications as advanced thermoelectric
materials~\cite{sales}. Recently, several groups have reported 
investigations
of a cerium based filled-skutterudite Kondo insulator,  
CeOs$_4$Sb$_{12}$. Here we make comparison of the theory
outlined in this paper to experimental measurements of d.c.\ and optical
transport in CeOs$_4$Sb$_{12}$.

\begin{figure}[h]
\epsfxsize=300pt
\begin{center}
\epsffile{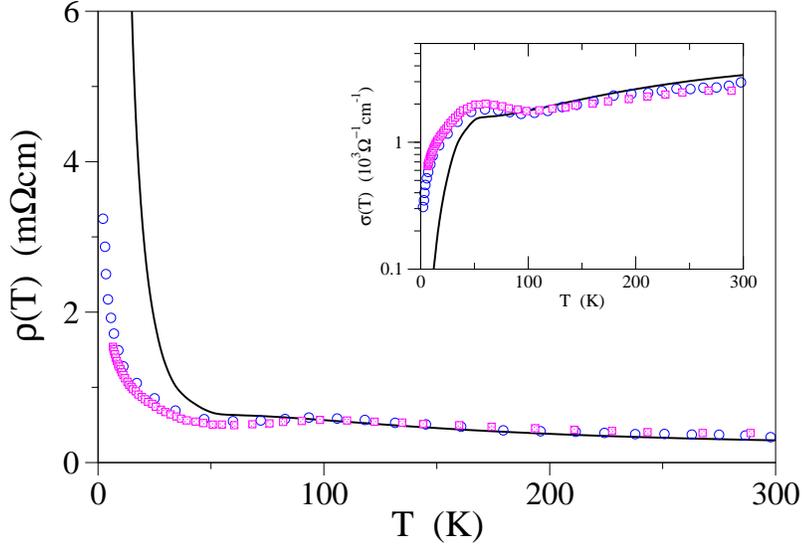}
\end{center}
\caption{\label{fig:fig6} Experimental d.c.\ resistivity of CeOs$_4$Sb$_{12}$ 
with the phonon contribution subtracted (circles~\cite{cosres} and
squares~\cite{cosres2}) compared to
the theoretically determined $\rho(T)$ (solid line), which has been rescaled
by $\om_L=86K$ and $1/\sigma_0=74\mu\Omega cm$. The inset shows the corresponding
d.c.\ conductivities. 
}
\end{figure}

We consider first the d.c.\ transport. In this compound the phonon 
contribution to the d.c.\ resistivity is quite significant, so for comparison to theory we need to extract the magnetic contribution, $\rho_{mag}(T)$, from the d.c.\ resistivity measured directly. This is achieved by subtracting the phonon contribution -- itself estimated conventionally as the resistivity of $LaOs_4Sb_{12}$~\cite{cosres2,laos} 
(see~\cite{vidh05-2} for a detailed discussion) -- from the experimentally measured d.c.\ resistivity~\cite{cosres}; and is shown as circles in figure~\ref{fig:fig6}. Another experimental group has also reported $\rho_{mag}(T)$ itself~\cite{cosres2}; and although
their absolute magnitudes are quite different from those of~\cite{cosres}, 
a simple y-axis rescaling is sufficient to collapse the two sets of data, as
indicated by squares and circles in figure~\ref{fig:fig6} 
(and indicating simply distinct sample geometries in the two cases).

\begin{figure}[t]
\begin{center}
\includegraphics[height=5cm,clip]{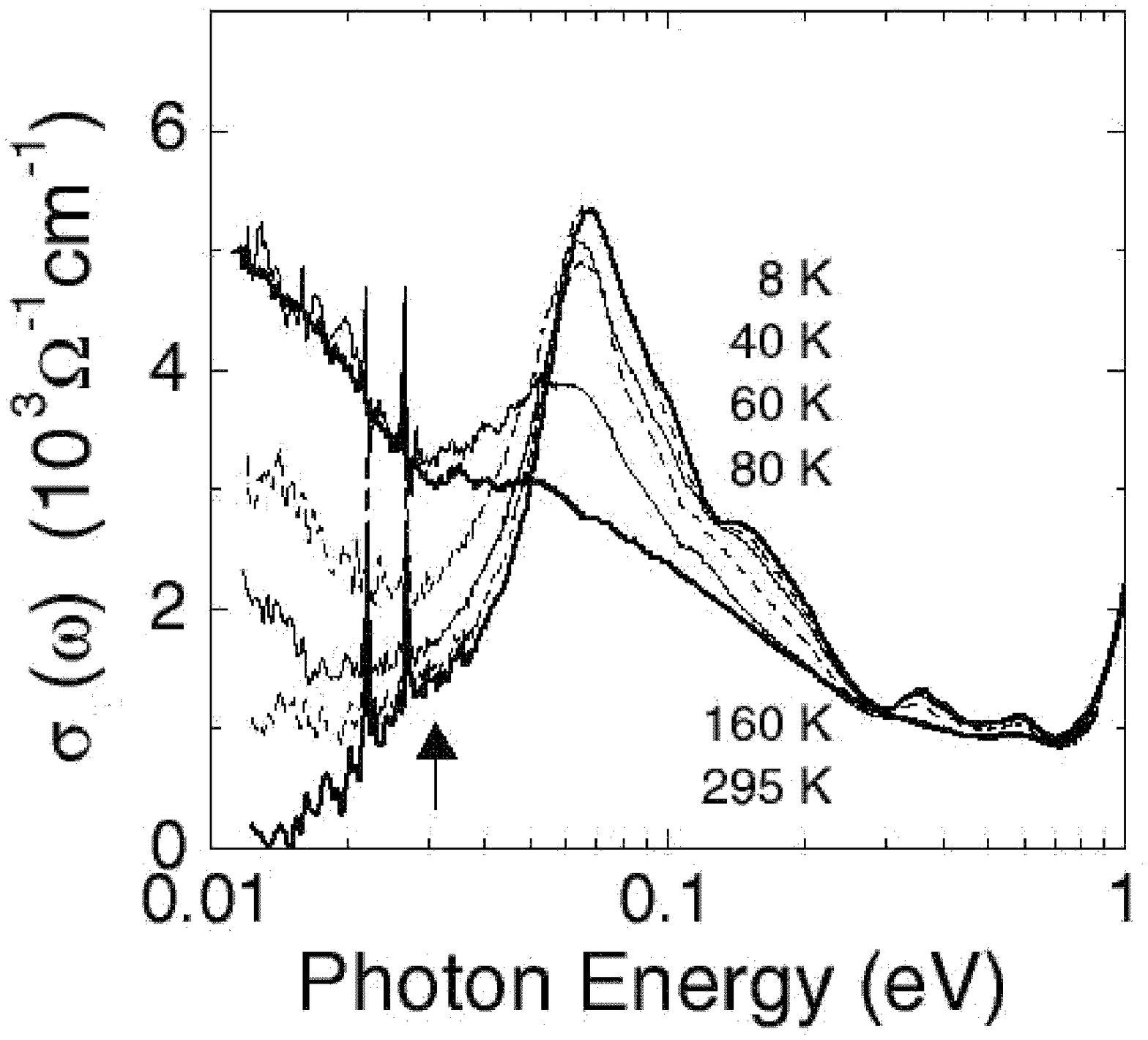}
\includegraphics[height=5cm,clip]{opt_comp.eps}
\end{center}
\caption{\label{fig:fig7} The left panel shows the experimentally measured optical
conductivity of CeOs$_4$Sb$_{12}$~\cite{cosopt} for various temperatures. The right panel shows the theoretically determined $\sigma(\om;T)$ 
for the same model parameters as in figure~\ref{fig:fig6} and similar temperatures
to experiment. 
}
\end{figure}

The low temperature ($T\lesssim 30K$) resistivity can be fitted to
a $\exp[(T^*/T)^{1/2}]$ form, suggesting the dominant mechanism
of transport in this temperature range to be Efros-Shklovskii type variable
range hopping~\cite{hedo} (which extrinsic behaviour is not of course 
included in the present theory). A shallow maximum is seen in figure~\ref{fig:fig6} at
$T\sim 100K$, beyond which $\rho_{mag}(T)$ decays monotonically.
A rough estimate of the low energy scale ($\om_L\equiv ZV^2/t_*$) in this 
system may be obtained through an estimation of the (indirect) gap
in the experimental optical conductivity~\cite{cosopt}, shown in
the left panel of figure~\ref{fig:fig7}. This yields $\Delta_{ind}\sim 15meV
\simeq 175K$, which thus gives the low energy scale $\om_L
=\Delta_{ind}/2\simeq 90K$ (see also the left panel of figure~\ref{fig:fig3}
and~\cite{vidh03}).

  An idea of the parameter regime this system belongs to can be gleaned from 
two qualitative features evident in the $(\om,T)$-dependence of the experimental $\sigma_{\exp}(\om;T)$~\cite{cosopt} shown in figure~\ref{fig:fig7}.
First, the highest temperature at which $\sigma_{\exp}(\om;T)$ is measured is $295K$ which is 
$\sim 3\om_L$, while the optical conductivity at the same temperature is affected up to
$\sim 0.3eV \simeq 40 \om_L$.  The fact that the thermally induced spectral weight
redistribution is over a range of frequencies much higher than the temperature at which 
$\sigma_{\exp}(\om;T)$ is measured is characteristic of relatively weak 
correlations~\cite{vidh03}. Second, a  closer look at the low-frequency structure of 
$\sigma(\om;T=8K)$ reveals an additional absorption feature at $\om \sim 30 meV$, which
in the context of the theoretical results obtained above (see e.g.\ figure~\ref{fig:fig3}), is suggestive of mixed-valence character. 

With this in mind, we choose an $\ep_c=-0.3$, and for various $U,V^2$ calculate
$\sigma(\om;T)$. We find that the  d.c.\ resistivity ($\om =0$) and the optical conductivity determined at  $U\sim 5.5, V^2=0.5$ match best with the experiment, with $\om_L\sim 86K$. 
With these parameter values the $f$-level asymmetry $\eta$ is found to be $0.9$, implying
$n_f=0.75$ and $n_c=1.25$ consistent with mixed valent behaviour, while the quasiparticle weight $Z\simeq0.083$ implying a relatively low effective mass  $\sim 12$ consistent with modest correlations. The resultant theoretical d.c.\ resistivity is shown in 
figure~\ref{fig:fig6} (solid line, with the $x$-axis scaled by $\om_L=86K$ and the $y$-axis by $1/\sigma_0=74\mu\Omega cm$). And comparison to experiment is seen to be rather good for $T\gtrsim 30K$ or so (recall as above that transport for $T\lesssim 30 K$ is dominated by variable range hopping~\cite{hedo}, which is naturally absent from the theory).

The theoretical optical conductivity evaluated for the same model
parameters as in figure~\ref{fig:fig6}  is shown in the right panel of figure~\ref{fig:fig7},
for similar temperatures as the experiment~\cite{cosopt} (left panel). 
An arrow marks the additional absorption seen as a shoulder, which arises at 
$\om \sim 40 meV$ and thus compares well with the experimental value of $\sim 30meV$. 
The theoretical direct gap peak appears at $\om \sim 100meV$, while the corresponding experimental peak position is $\sim 70meV$.  The overall theoretical lineshape and its thermal evolution also matches rather well with experiment.  We add further that this comparison does not depend crucially on $\ep_c$ being equal to $-0.3$. Varying $\ep_c$ 
by $\pm 0.1$ does not change the qualitative picture, although optimal quantitative 
agreement arises for $\ep_c=-0.3, U=5.5$ and $V^2=0.5$ as employed above.

\section{Conclusion}

      We have described in this paper a many-body theory for mixed-valent Kondo 
insulators, employing a local moment approach to the periodic Anderson model 
within the framework of dynamical mean field theory. Kondo insulators were argued 
as a rule to be mixed-valent, with $n_f=2-n_c\neq 1$, and which regime of behaviour 
is not captured by the particle-hole symmetric limit of the PAM ($n_{f}=1=n_{c}$). 
To that end we have considered the general asymmetric PAM, together with the 
constraint $n_f+n_c=2$ which ensures an insulating gap in the ($T=0$) 
single-particle spectrum and related dynamics.

  Single-particle spectra, as well as optical and d.c.\ conductivities, have been considered, and exhibit features specific to mixed-valent behaviour. The $\mathbf{k}$-resolved conduction electron spectra are found to contain additional absorption features above the Fermi level, giving rise in turn to an intrinsic shoulder-like absorption in the optical conductivity at frequencies lower than the direct (mid-infrared) gap. We emphasise that such a structure is found to be characteristic of the {\em interacting mixed-valent} case, and not restricted to either an insulating or metallic ground state.Thus we believe that the shoulder-like feature seen experimentally in the optical conductivity of mixed-valent materials such as the insulator CeOs$_4$Sb$_{12}$ and the metal YbAl$_3$, arises intrinsically from a combination 
of many-body scattering and intermediate valence. A direct comparison between theory and transport/optical experiments on CeOs$_4$Sb$_{12}$ has been given. This yields good quantitative agreement, both reaffirming the view of CeOs$_4$Sb$_{12}$ as a mixed-valent hybridization gap material~\cite{cosres,hedo} and showing that the theory described here for
generic mixed-valent Kondo insulators can account for the transport/optical properties of these systems. \\

\bf Acknowledgements. \rm We are grateful to E. D. Bauer for providing us with his d.c.\ resistivity data for LaOs$_{4}$Sb$_{12}$, and to the EPSRC for supporting this research.

\end{document}